\documentclass{article}
\usepackage{spconf,amsmath,graphicx}
\usepackage{xcolor}
\usepackage{multirow}
\usepackage{cite}
\usepackage{gensymb}
\usepackage{amsfonts}
\usepackage{bm}

\usepackage{amssymb}
\usepackage{pifont}
\newcommand{\cmark}{\ding{51}}%
\newcommand{\xmark}{\ding{55}}%

\title{L-SpEx: Localized Target Speaker Extraction}
\name{Meng Ge$^{1,2}$, Chenglin Xu$^{3,*}$, Longbiao Wang$^{1, *}$, Eng Siong Chng$^4$, Jianwu Dang$^{1, 5}$, Haizhou Li$^{2,6}$ 
\thanks{This work is supported by A*STAR under its RIE2020 Advanced Manufacturing and Engineering Domain (AME) Programmatic Grant (Grant No. A1687b0033); National Research Foundation, Singapore under its AI Singapore Programme (AISG Award No: AISG-100E-2018-006); Human-Robot Interaction Phase 1 (Grant No. 192 25 00054), National Research Foundation (NRF) Singapore under the National Robotics Programme; National Natural Science Foundation (61771333), Tianjin Municipal Science and Technology Project (18ZXZNGX00330).
This research is also funded by the Deutsche Forschungsgemeinschaft (DFG, German Research Foundation) under Germany's Excellence Strategy (University Allowance, EXC 2077,
University of Bremen).
$^*$ Corresponding author.}}

\address{
  $^1$ College of Intelligence and Computing, Tianjin University, Tianjin, China\\
  $^2$ Department of Electrical and Computer Engineering, National University of Singapore, Singapore \\
  $^3$ Kuaishou Technology, Beijing, China \\
  $^4$ School of Computer Science and Engineering, Nanyang Technological University, Singapore\\
  $^5$ Japan Advanced Institute of Science and Technology, Ishikawa, Japan \\
  $^6$ The Chinese University of Hong Kong, Shenzhen, China}
  
\begin{document}
%
\maketitle
\begin{abstract}
Speaker extraction aims to extract the target speaker's voice from a multi-talker speech mixture given an auxiliary reference utterance. 
Recent studies show that speaker extraction benefits from the location or direction of the target speaker.
However, these studies assume that the target speaker's location is known in advance or detected by an extra visual cue, e.g., face image or video. In this paper, we propose an end-to-end localized target speaker extraction on pure speech cues, that is called L-SpEx. Specifically, we design a speaker localizer driven by the target speaker's embedding to extract the spatial features, including direction-of-arrival (DOA) of the target speaker and beamforming output. Then, the spatial cues and target speaker's embedding are both used to form a top-down auditory attention to the target speaker.
Experiments on the multi-channel reverberant dataset called MC-Libri2Mix show that our L-SpEx approach significantly outperforms the baseline system.
\end{abstract}
\section{Introduction}
\label{sec:intro}
Human has the ability to selectively listen to a particular speaker through various stimuli in a multi-talker scenario, that is called selective auditory attention in \textit{cocktail party} problem \cite{cherry1953some}. Ever since the theory is proposed, researchers never stop seeking the engineering solution to confer human's selective attention capability on machines, as there is high demand for various real-world applications, such as speech recognition \cite{chen2020improved,shi2020sequence} and speaker verification \cite{rao2019target,xu2021target}.

With the advent of deep learning in recent years, blind speech separation methods have been widely studied to solve the cocktail party problem by applying neural networks \cite{hershey2016deep,yu2017permutation,luo2019conv,luo2020dual} and beamforming \cite{wu2020end,ochiai2020beam,chang2019mimo,zhang2020end}. The neural network seeks the regular patterns (i.e., masks) between the time-frequency representation of the target speech and mixture speech, while beamforming incorporates the spatial statistics (i.e., spatial covariance matrix) obtained from the estimated masks to compute beamformer's weights and filter the desired voice. However, speech separation always requires that the number of speakers is known as a prior, and assumes the label permutation is unchanged during training, which greatly limits its scope of real-world applications. 

Unlike blind speech separation, speaker extraction only extracts the target speech from a mixture speech driven by spectral \cite{spex2020,spex_plus2020} or spatial cues \cite{chen2018multi,gu2019neural,gu2020multi} of the target speaker. The spectral cue is always represented by a speaker embedding from an enrolled reference utterance, while the spatial cue is usually transformed into spectrum-like features derived from the target speaker's location. For example, Chen et al. \cite{chen2018multi} introduced a location-based angle feature to guide separation network, which was the cosine distance between the steering vector and inter-channel phase difference (IPD) for each speaker in the mixture. Gu et al. \cite{gu2019neural} suggested that the beamforming output could be regarded as an alternative way of spatial cues, as beamforming aimed to summarize the signals from the target speaker's direction and suppress non-target signals. However, these studies often require the speaker location is known in advance or detected using an extra visual cue.

To address this issue, we propose an end-to-end localized target speaker extraction on pure speech cues, that is called L-SpEx. We design a target speaker localizer driven by an enrolled utterance of the target speaker to extract the target speaker's DOA and beamforming output, simultaneously. By doing so, the extracted spatial cues and the enrolled utterance can be further used to guide the network to learn which direction and speaker to be extracted.

\begin{figure}[t]
	\centering
	\includegraphics[width=0.98\linewidth]{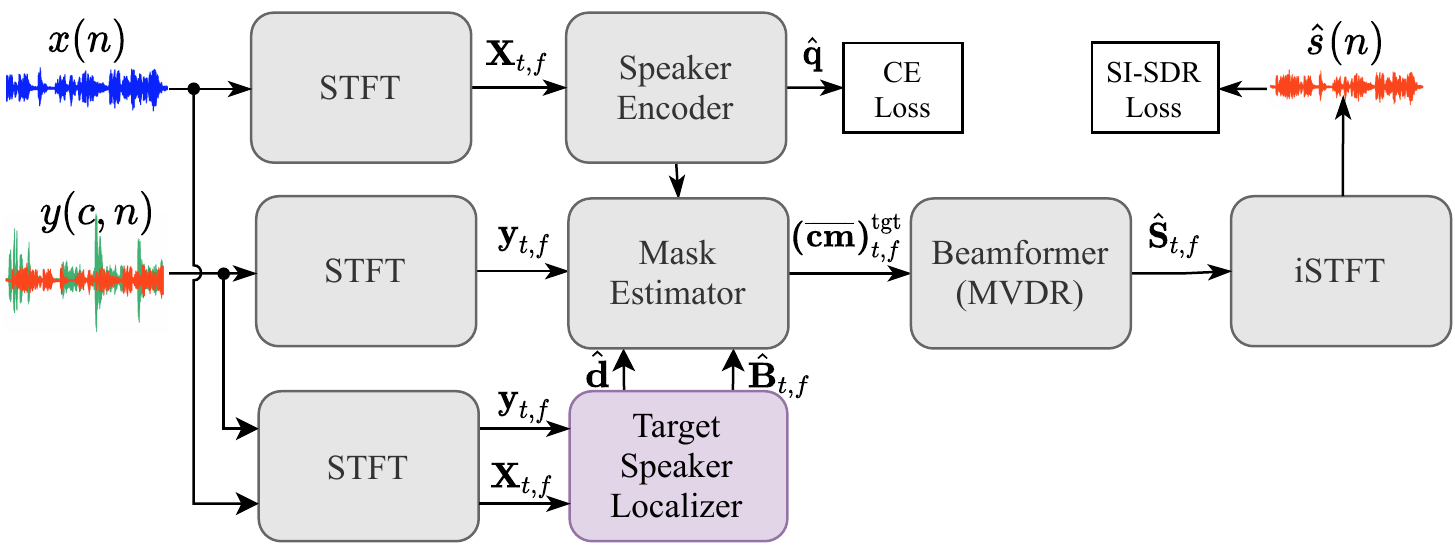}
	\caption{The diagram of the proposed L-SpEx. The target speaker localizer is illustrated in Fig. 2.}
	\label{fig:overall_framework}
	\vspace{-10pt}
\end{figure}

\section{L-SpEx Architecture}
\label{sec:format}

As illustrated in Fig. \ref{fig:overall_framework}, the  difference between the proposed L-SpEx system and other speaker extraction systems lies in the target speaker localizer (Fig. \ref{fig:target_doa_net}).


\subsection{Target speaker localizer}
\label{sec:target_speaker_localizer}
The target speaker localizer learns to encode the spatial cues related to the target speaker's direction from the multi-channel mixture signal $y(c,n)$, with reference to a reference utterance $x(n)$ by the target speaker. We have,
\begin{equation}
    y(c,n) = s(c,n) + \sum \nolimits_{i=1}^{I} b_i(c,n)
    \label{signal_definition}
\end{equation}
where $c$ denotes the channel index and $n$ denotes discrete time index. $s(c,n)$ represents the target signal and $b_i(c,n)$ represents the interference signal corresponding to speaker $i$.

\begin{figure}[t]
	\centering
	\includegraphics[width=0.8\linewidth]{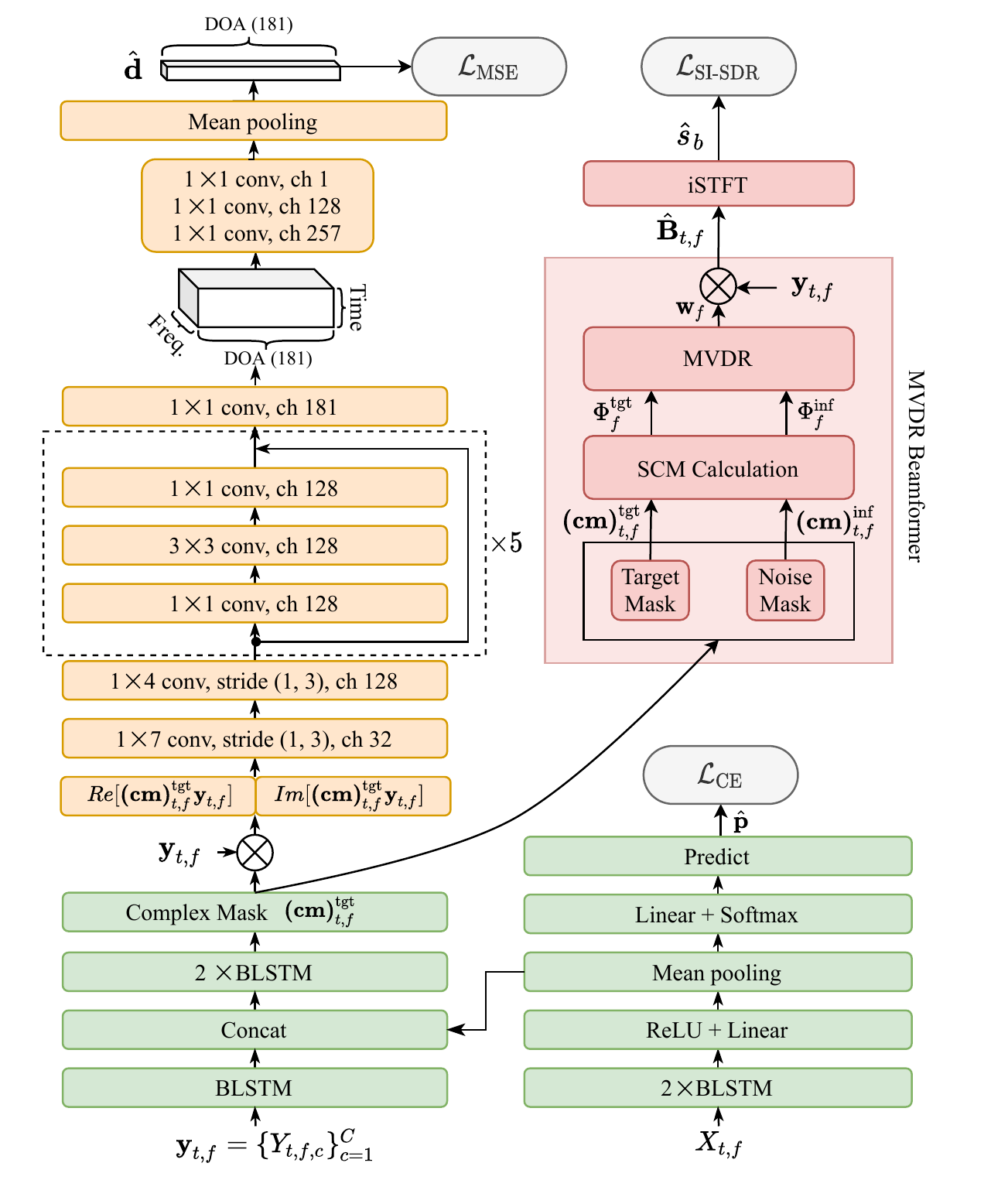}
	\caption{The network structure of target speaker localizer.}
	\label{fig:target_doa_net}
	\vspace{-10pt}
\end{figure}

Formally, let ${\bf{Y}}_{t,f,c} \in \mathbb{C}$ be the STFT coefficient of the $c$-th channel mixture signal $y(c,n)$ at time-frequency bin $(t,f)$, and let ${\bf{X}}_{t,f} \in \mathbb{C}$ be the STFT coefficient of an enrolled single-channel utterance $x(n)$ of the corresponding target speaker. As shown in Fig. \ref{fig:target_doa_net}, we first employ a network to estimate a complex-valued mask, as opposed to a real-valued mask, for speaker location estimation. This is motivated by  Sharath's work \cite{adavanne2018sound}, which shows that speaker location estimation relies strongly on both phase-differences and magnitude-differences between the microphones. Thus, the complex-valued masks ${\bf{(cm)}}_{t,f}^{\text{tgt}}$ is estimated as follow:
\begin{equation}
\begin{aligned}
    {\bf{(cm)}}_{t,f}^{\text{tgt}} &= \{ Re[{\bf{(cm)}}_{t,f}^{\text{tgt}}], Im[{\bf{(cm)}}_{t,f}^{\text{tgt}}] \} \\
    &= \text{CMaskEst}({\bf{y}}_{t,f}, \text{Enc}_{\text{speaker}}({\bf{X}}_{t,f}))
\end{aligned}
\end{equation}
where ${\bf{y}}_{t,f} = \{{\bf{Y}}_{t,f,c} \}_{c=1}^C \in \mathbb{C}^C$ is the spatial vector of the signals obtained from all $C$-microphones for each time-frequency bin $(t,f)$. $\text{CMaskEst}(\cdot)$ and $\text{Enc}_{\text{speaker}}(\cdot)$ represent a complex mask estimator and speaker encoder, and both structures are built with several BLSTM layers as shown in Fig. \ref{fig:target_doa_net}. $Re[\cdot]$ and $Im[\cdot]$ denotes the real and imaginary part of a complex tensor, respectively. 

After the masks are estimated, we use ${\bf{(cm)}}_{t,f}^{\text{tgt}} {\bf{y}}_{t,f}$ as the input of DOA estimator to predict target speaker's DOA. We employ two CNN layers with kernel size of $1 \times 7$ and $1 \times 4$ on the masked input, followed by residual network blocks with a number of 5. Then, a $1 \times 1$ convolutional layer with 181 output channels (corresponding to 181 azimuth directions) projects the features to the DOA space. Finally, three $1 \times 1$ convolutional layers and a mean pooling operation summaries the time and frequency to obtain the 181-dimension DOA vector:
\begin{equation}
    \hat{{\bf{d}}} = \text{DOAEst}({\bf{(cm)}}_{t,f}^{\text{tgt}} {\bf{y}}_{t,f})
\end{equation}
where $\text{DOAEst}(\cdot)$ denotes the DOA estimator.

Motivated by the studies \cite{gu2019neural,wang2018combining}, beamforming output is also regarded as a direction-related spatial cue, as beamforming has the ability to summarize the signals from target speaker's direction. Therefore, the estimated masks are also used to compute the cross-channel spatial covariance matrices (SCMs) ${\bf{\Phi}}_{f}^{j}$ and then obtain the beamforming output $\hat{{\bf{B}}}_{t,f}$,
\begin{gather}
    {\bf{\Phi}}_{f}^{j} = \frac{1}{\sum_{t=1}^{T} {\bf{(cm)}}_{t,f}^{j}} \sum\nolimits_{t=1}^{T} {\bf{(cm)}}_{t,f}^{j} {\bf{y}}_{t,f} {\bf{y}}_{t,f}^{H}
    \label{eq:scm_calculation} \\
    \hat{{\bf{B}}}_{t,f} = {\bf{w}}_{f}^{H} {\bf{y}}_{t,f} = \left [\frac{({\bf{\Phi}}_{f}^{\text{inf}})^{-1} {\bf{\Phi}}_{f}^{\text{tgt}}}{\text{tr}(({\bf{\Phi}}_{f}^{\text{inf}})^{-1} {\bf{\Phi}}_{f}^{\text{tgt}})} {\bf{u}} \right]^{H} {\bf{y}}_{t,f}
    \label{eq:mvdr_weight_calculation}
\end{gather}
Here, $j \in \{\text{tgt}, \text{inf}\}$ and ${\bf{m}}_{t,f}^{\text{inf}} = 1 - {\bf{m}}_{t,f}^{\text{tgt}}$. ${\bf{u}} \in \mathbb{R}^{C}$ is a vector denoting the reference microphone, and $\text{tr}(\cdot)$ denotes the trace operation. ${\bf{w}}_{f} = \{{\bf{W}}_{t,f,c} \}_{c=1}^C \in \mathbb{C}^C$ is corresponding to time-invariant beamformer coefficients. $C$ denotes the number of channels and $H$ denotes the conjugate transpose.

\subsection{Localized target speaker extraction}
\label{sec:pagestyle}

Given the estimated DOA likelihood coding $\hat{{\bf{d}}}$ and beamforming output $\hat{{\bf{B}}}_{t,f}$ of the target speaker in Sec. \ref{sec:target_speaker_localizer}, we can obtain two direction-related spatial features, i.e., $\text{DF}_{\text{angle}}$ and $\text{DF}_{\text{beam}}$. The former \cite{chen2018multi} is derived from the estimated target speaker's angle $\hat{\theta}=\text{argmax}(\hat{{\bf{d}}})$, the latter is derived from beamforming output $\hat{{\bf{B}}}_{t,f}$, which are defined as follows
\begin{gather}
    \text{DF}_{\text{angle}}(t,f) = \frac{1}{P} \sum_{l,r \in \Omega} cos({\bf{o}}_{l,r}-\frac{\pi f_s f \Delta_{l,r} cos \hat{\theta}}{(N_{\text{FFT}}-1)v}), \label{eq:angle_df_calculation} \\
    \text{DF}_{\text{beam}}(t,f) = \sqrt{Re[\hat{{\bf{B}}}_{t,f}]^2 + Im[\hat{{\bf{B}}}_{t,f}]^2}
    \label{eq:beam_df_calculation}
\end{gather}
where $\Omega$ contains $P$ microphone pairs, and ${\bf{o}}_{l,r}=\angle{{\bf{Y}}_{t,f,l}}-\angle{{\bf{Y}}_{t,f,r}}$ represents the observed inter-channel phase difference (IPD) between left channel $l$ and right channel $r$. $N_{\text{FFT}}$ is the number of FFT bins, $v$ is the sound velocity and $f_s$ is the sampling rate. Note that $f$ ranges from $0$ to $(N_{\text{FFT}}-1)$. $\Delta_{l,r}$ denotes the distance between the microphone pair $(l,r)$.

The above two direction-related spatial features have an ability to inform the extraction network of target speaker's direction, while the speaker embedding obtained from a speaker encoder can guide the network to attend to the target speaker. Thus, we use both spatial features as the inputs of the mask estimator to predict better masks for the target speaker extraction. The better masks ${\bf{(\overline{cm})}}_{t,f}^{\text{tgt}}$ are calculated as
\begin{gather}
{\bf{y}}_{t,f}^{\text{new}} = \text{Concat}[{\bf{y}}_{t,f},\text{DF}_{\text{beam}},\text{DF}_{\text{angle}}], \\
{\bf{(\overline{cm})}}_{t,f}^{\text{tgt}}=\overline{\text{CMaskEst}}\{{\bf{y}}_{t,f}^{\text{new}}, \overline{\text{Enc}_{\text{speaker}}}({\bf{X}}_{t,f})\}
\end{gather}
where $\text{Concat}[\cdot]$ is the concatenation operation. $\overline{\text{CMaskEst}}$ and $\overline{\text{Enc}_{\text{speaker}}}$ denote the complex mask estimator and speaker encoder in extraction network for speaker extraction.

Finally, the masks are used to obtain the extracted STFT spectrum $\hat{\bf{S}}_{t,f} \in \mathbb{C}$ by using MVDR formula, i.e., Eq. (\ref{eq:scm_calculation}) and (\ref{eq:mvdr_weight_calculation}), and further estimate the target signal $\hat{s}$ via $\text{iSTFT}$,
\begin{equation}
\hat{s} = \text{iSTFT}(\hat{\bf{S}}_{t,f}) = \text{iSTFT}(\text{MVDR}({\bf{(\overline{cm})}}_{t,f}^{\text{tgt}}, {\bf{y}}_{t,f}))
\label{eq:signal_extraction}
\end{equation}

\subsection{End-to-end training}

We first pretrain the target speaker localizer by a multi-task learning strategy, and then optimize the whole network. The loss function of target speaker localizer is defined as follow:
\begin{gather}
\mathcal{L}_{\text{localizer}} = \mathcal{L}_{\text{SI-SDR}}(\hat{s_b}, s) +
\alpha \mathcal{L}_{\text{CE}}(\hat{{\bf{p}}}, {\bf{p}}) + \beta \mathcal{L}_{\text{MSE}}(\hat{{\bf{d}}}, {\bf{d}}) \nonumber\\
\mathcal{L}_{\text{SI-SDR}}(\hat{s_b}, s) = -20 \log_{10}\frac{||(\hat{s_b}^T s / s^T s) \cdot s||}{||(\hat{s_b}^T s / s^T s) \cdot s - \hat{s_b}||} \nonumber\\
\mathcal{L}_{\text{CE}}(\hat{{\bf{p}}}, {\bf{p}}) = -\sum\nolimits_{i=1}^{N} {\bf{p}}_i \log(\hat{\bf{p}}_i) \nonumber\\
\mathcal{L}_{\text{MSE}}(\hat{{\bf{d}}}, {\bf{d}}) = \sum\nolimits_{i=1}^{M} ||\hat{{\bf{d}}}_i - {\bf{d}}_i|| \nonumber
\label{eq:speaker_localizer_loss}
\end{gather}
where $\alpha$ and $\beta$ are scaling factors. $\mathcal{L}_{\text{SI-SDR}}$ aims to minimize the signal reconstruction error. $\hat{s_b}=\text{iSTFT}(\hat{{\bf{B}}}_{t,f})$ and $s$ are the estimated signal and the target clean signal of reference microphone, respectively. $\mathcal{L}_{\text{CE}}$ is the cross-entropy loss for speaker classification. $N$ denotes the number of speakers. ${\bf{p}}_i$ is the true class label for speaker $i$, and $\hat{\bf{p}}_i$ represents the predicted probability corresponding to $i$-th speaker. $\mathcal{L}_{\text{MSE}}$ is the mean squared error for target DOA estimation. $M$ is the number of azimuth directions, here $M=181$. $\hat{{\bf{d}}}_i$ and ${\bf{d}}_i$ are the predicted and ground-truth DOA coding of the target speaker. Based on the likelihood-based coding in \cite{he2021neural}, the desired ground-truth values ${\bf{d}}_i$ are defined as follows:
\begin{equation}
{\bf{d}}_i=
\begin{cases}
 e^{-d(\theta_i,\theta)^2/\sigma^2}, & \mbox{if $\theta$ exists}\\
0, & \mbox{otherwise}
\end{cases}
\end{equation}
where $\theta$ is true target speaker's angle and $\theta_i$ is one of 181 azimuth directions. $\sigma$ is the parameter to control the width of the Gaussian curves. $d(\cdot, \cdot)$ is the azimuth angular distance. 

After target speaker localizer is pretrained, we optimize the whole L-SpEx network by using the following loss,
\begin{equation}
\mathcal{L}_{\text{L-SpEx}} = \mathcal{L}_{\text{SI-SDR}}(\hat{s}, s) +
\gamma \mathcal{L}_{\text{CE}}(\hat{{\bf{q}}}, {\bf{p}}),
\end{equation}
where $\hat{s}$ is the extracted signal via Eq. \ref{eq:signal_extraction}, and $\hat{\bf{q}}$ is predicted probability from the speaker encoder $\overline{\text{Enc}_{\text{speaker}}}$. $\gamma$ is also a scale factor to balance the two objectives.


\vspace{-5pt}
\section{Experiments and Discussion}
\label{sec:typestyle}
\vspace{-5pt}

To facilitate the evaluation, we introduce a multi-channel reverberated version of the Libri2Mix\footnote{https://github.com/JorisCos/LibriMix} dataset which we refer to as MC-Libri2Mix. The original Libri2Mix is a clean 2-talker mixture corpus generated from the LibriSpeech corpus by mixing two randomly selected utterances. Libri2Mix contains training data with 64,700 utterances (270 hours, 1,172 speakers), development data with 3,000 utterances (11 hours, 40 speakers), and test data with 3,000 utterances (11 hours, 40 speakers). The average duration of the utterances is 14.8s. 

The room impulse responses (RIRs) in MC-Libri2Mix is simulated using pyroomacoustics\footnote{https://github.com/LCAV/pyroomacoustics} package. For the room configurations, the length and width of each room are randomly drawn in the range $[5, 10]$m, and the height is selected in the range $[3, 4]$m. The reverberation time ($\text{RT}_{60}$) of the reverberant data ranges from 200ms to 600ms. In MC-Libri2Mix,  we consider a linear array with four microphones, where the microphone-to-microphone distance is 5cm~\cite{wang2018combining}. Target speakers are placed in the frontal plane and are at least $15^{\circ}$ apart from each other. The distance of speaker and the center of microphone is from 0.75m to 2m.

Unlike in blind speech separation, we set that the speakers in each 2-talker mixed speech acted as the target speaker in turn, and the corresponding auxiliary reference speech is randomly selected from original LibriSpeech corpus. In practice, the training set (127,056 examples, 1,172 speakers) and development set (2,344 examples, 1,172 speakers) are randomly selected from the training data of MC-Libri2Mix. The test set contains 6,000 examples from the test data of MC-Libri2Mix. The details of configuration and data simulation can be found at https://github.com/gemengtju/L-SpEx.git

\vspace{-5pt}
\subsection{Experimental setup}
\vspace{-5pt}

We train all systems for 70 epochs on the 4-channel mixture segments and their corresponding reference utterances. The learning rate is initialized to $1e^{-4}$ and decays by 0.5 if the accuracy of validation set is not improved in 2 consecutive epochs. Early stopping is applied if no best model is found in the validation set for 5 consecutive epochs. Adam is used as the optimizer. For feature extraction, STFT is performed with a 8k Hz sampling rate and a 25ms window length with a 10ms stride, and the feature dimension is 257. For mask estimation network, we use three BLSTM layers with 512 cells, and the dimension of speaker embedding is 256. The speaker embedding is inserted between first and second BLSTM layer. For the loss configuration, we used $\alpha=\gamma=0.5, \beta=10$ to balance the loss. The parameter $\sigma$ in Gaussian curve is 6. 

	

\renewcommand{\arraystretch}{1.5}
\begin{table}[tp]
	
	\centering
	\fontsize{5.4}{5}\selectfont
	\caption{SDR (dB) and SI-SDR (dB) in a comparative study on the MC-Libri2Mix dataset under open condition. ``${\bf{m}}$'' and ``${\bf{cm}}$'' represent the real-value mask and complex-value mask, respectively. ``Pretrained Speaker Localizer" indicates the extracted signal $\hat{s}_b$ that is derived from the output spectrum $\hat{{\bf{B}}}_{t,f}$ of the pretrained target speaker localizer in Fig. \ref{fig:target_doa_net}.}
	\begin{tabular}{|c|c|c|c|c|c|c|c|}
		\hline
		\multirow{2}{*}{ID}
		&\multirow{2}{*}{Methods}
		&\multirow{2}{*}{\shortstack{Mask\\Type}} &\multicolumn{2}{c|}{Spatial Cues}
		&\multirow{2}{*}{\shortstack{E2E\\Train}}
		&\multirow{2}{*}{SDR} &\multirow{2}{*}{SI-SDR} \\\cline{4-5}
		& & &$\text{DF}_{\text{beam}}$ &$\text{DF}_{\text{angle}}$ & & & \cr
		\hline
		\hline
		1 &Unprocessed &- &- &- &- &0.46 &0.07 \\
		2 &Mask MVDR (${\bf{m}}$) &${\bf{m}}$ &\xmark &\xmark &- &8.03 &6.36 \\
		3 &Mask MVDR (${\bf{cm}}$) &${\bf{cm}}$ &\xmark &\xmark &- &8.02 &6.26 \\
		4 &Pretrained Speaker Localizer &${\bf{cm}}$ &\xmark &\xmark &- &7.44 &5.80 \\\hline
		5 &\multirow{3}{*}{L-SpEx} &${\bf{cm}}$ &\cmark &\xmark &\xmark &8.96 &7.17 \\
		6 & &${\bf{cm}}$ &\cmark &\cmark &\xmark &9.41 &7.29 \\
		7 & &${\bf{cm}}$ &\cmark &\cmark &\cmark &\textbf{9.68} &\textbf{7.45} \cr
		\hline
	\end{tabular} \label{tbl:cmp_mc_libr2imix}
	\vspace{-10pt}
\end{table}

\vspace{-5pt}
\subsection{Results and analysis}
\vspace{-5pt}

We compare L-SpEx with the mask-based MVDR baseline systems on MC-Libri2Mix in terms of SDR and SI-SDR. From Table \ref{tbl:cmp_mc_libr2imix}, we conclude: 1) Our L-SpEx approach achieves the best performance under the open condition. Compared to the ``Mask MVDR (${\bf{cm}}$)'' system, L-SpEx leads to 20.7\% and 19.0\% relative improvement in terms of SDR and SI-SDR measure, respectively. The improvements mainly come from the augmented spatial features through the proposed target speaker localizer module. 2) The proposed L-SpEx with $\text{DF}_{\text{beam}}$ achieves 0.94 dB and 0.89 dB performance gain over the ``Mask MVDR (${\bf{cm}}$)'' baseline in terms of SDR and SI-SDR. This result proves that the spatial feature derived from beamforming outputs can let network attend to the signal from the target direction and ignore non-target signals. 3) The result of L-SpEx system with $\text{DF}_{\text{beam}}$ and $\text{DF}_{\text{angle}}$ further show the effectiveness of spatial feature. This shows that $\text{DF}_{\text{angle}}$ and $\text{DF}_{\text{beam}}$ represents different aspects of spatial cues of target speaker, and the two features complement each other. 4) The result evaluated on the output of pre-trained speaker localizer is worse than the mask-based MVDR systems. The reason is that the DOA estimation and beamforming have something in common, but they are still two separate tasks and one mask is hard to optimize.

We further report the speaker extraction performance on different angle distance mixture speech in Table \ref{tbl:cmp_mc_librimix_diff_angle}. From Table \ref{tbl:cmp_mc_librimix_diff_angle}, we find that the mixture speech with smaller angle difference (i.e., $<45^{\circ}$) is more difficult to extract target speaker. It is observed that our L-SpEx system achieves the SI-SDR from 5.82 dB to 7.06 dB, which is even higher than the performance of ``Mask MVDR (${\bf{cm}}$)'' on large angle distance (i.e., $>45^{\circ}$). Furthermore, we report the extraction performance with different and same gender mixture speech, separately in Table \ref{tbl:cmp_mc_librimix_gender}. We observe that separating same-gender mixture is a more challenging task, as the spectral cues between each speaker are similar. When we incorporate the direction-related spatial information (i.e., $\text{DF}_{\text{beam}}$ and $\text{DF}_{\text{angle}}$) to increase the discrimination between speakers, the extraction process starts to be easier. Specifically, our L-SpEx system improve 1.19 dB SI-SDR performance compared with the spectral-only ``Mask MVDR (${\bf{cm}}$)'' system.

\renewcommand{\arraystretch}{1.5}
\begin{table}[tp]
	
	\centering
	\fontsize{5.4}{5}\selectfont
	\caption{SDR (dB) and SI-SDR (dB) in a comparative study of different angle distance under open condition. The percentage in the table head indicates the ratio of each condition.}
	\begin{tabular}{|c|c|c|c|c|c|c|c|c|}
		\hline
		\multirow{3}{*}{ID}
		&\multirow{3}{*}{Methods}
		&\multicolumn{2}{c|}{$<45^{\circ}$} &\multicolumn{2}{c|}{$45^{\circ} \text{-} 90^{\circ}$}  &\multicolumn{2}{c|}{$>90^{\circ}$} \\
		& &\multicolumn{2}{c|}{(34.6\%)} &\multicolumn{2}{c|}{(36.5\%)}  &\multicolumn{2}{c|}{(28.9\%)} \\ \cline{3-8}
		& &SDR &SI-SDR &SDR &SI-SDR &SDR &SI-SDR \cr
		\hline
		\hline
		1 &Unprocessed &0.46 &0.06 &0.43 &0.06 &0.48 &0.08 \\
		2 &Mask MVDR (${\bf{m}}$) &7.52 &5.92 &8.35 &6.66 &8.23 &6.51 \\
		3 &Mask MVDR (${\bf{cm}}$) &7.49 &5.82 &8.37 &6.57 &8.22 &6.42 \\
		4 &Pretrained Speaker Localizer &6.95 &5.36 &7.72 &6.09 &7.65 &6.00 \\\hline
		5 &\multirow{3}{*}{L-SpEx} &8.27 &6.55 &9.37 &7.56 &9.26 &7.45 \\
		6 & &8.64 &6.63 &\textbf{9.88} &\textbf{7.68} &9.75 &7.59 \\
		7 & &\textbf{8.97} &\textbf{7.06} &9.78 &7.66 &\textbf{9.75} &\textbf{7.66} \cr
		\hline
	\end{tabular} \label{tbl:cmp_mc_librimix_diff_angle}
	\vspace{-10pt}
\end{table}

\renewcommand{\arraystretch}{1.5}
\begin{table}[tp]
	
	\centering
	\fontsize{6.5}{5}\selectfont
	\caption{SDR (dB) and SI-SDR (dB) in a comparative study of different and same gender mixture under open condition.}
	\begin{tabular}{|c|c|c|c|c|c|}
		\hline
		\multirow{2}{*}{ID}
		&\multirow{2}{*}{Methods}
		&\multicolumn{2}{c|}{Diff. Gender (25.2\%)} &\multicolumn{2}{c|}{Same Gender (74.8\%)}\\ \cline{3-6}
		& &SDR &SI-SDR &SDR &SI-SDR\cr
		\hline
		\hline
		1 &Unprocessed &0.35 &-0.05 &0.49 &0.11 \\
		2 &Mask MVDR (${\bf{m}}$) &9.91 &8.22 &7.39 &5.74 \\
		3 &Mask MVDR (${\bf{cm}}$) &9.82 &8.02 &7.42 &5.67 \\
		4 &Pretrained Speaker Localizer &9.02 &7.39 &6.90 &5.28 \\\hline
		5 &\multirow{3}{*}{L-SpEx} &10.45 &8.76 &8.46 &6.64 \\
		6 & &11.10 &9.00 &8.85 &6.71  \\
		7 & &\textbf{11.11} &\textbf{9.20} &\textbf{8.94} &\textbf{6.86} \cr
		\hline
	\end{tabular} \label{tbl:cmp_mc_librimix_gender}
	\vspace{-10pt}
\end{table}

\section{Conclusions}

In this paper, we proposed an end-to-end localized target speaker extraction on pure speech cues, that is called L-SpEx, to eliminate the assumption of known target angle in existing studies. We took the advantages of an target speaker's enrolled utterance to design speaker localizer for estimating direction-related spatial cues. Experiments showed that the extracted spatial cues and the enrolled utterance input let extraction network works better, as target speech can be extracted based on speaker and direction aspects.



\vfill\pagebreak

\bibliographystyle{IEEEbib}
\bibliography{strings,refs}

\end{document}